\documentclass[american,aps, pre, reprint, showpacs, superscriptaddress]{revtex4-1}
\usepackage[T1]{fontenc}
\usepackage[utf8]{inputenc}
\usepackage{color}
\usepackage{babel}
\usepackage{array}
\usepackage{textcomp}
\usepackage{amsmath}
\usepackage{amssymb}
\usepackage{graphicx}
\usepackage[unicode=true,
 bookmarks=true,bookmarksnumbered=false,bookmarksopen=false,
 breaklinks=false,pdfborder={0 0 0},pdfborderstyle={},backref=false,colorlinks=true]
 {hyperref}
\hypersetup{
 colorlinks=true,linkcolor=blue,citecolor=blue}
\usepackage{comment}
\usepackage{lmodern}

\makeatletter

\providecommand{\tabularnewline}{\\}

\makeatother

\begin{document}

\title{Localized Faraday patterns under heterogeneous parametric excitation}

\author{Héctor Urra}
\altaffiliation{Current address: Sorbonne Université, Laboratoire PMMH – ESPCI Paris, 10 rue Vauquelin, 75005, Paris, France}
\affiliation{Instituto de F\'{i}sica, Pontificia Universidad Católica de Valpara\'{i}so,
Casilla 4059, Chile}

\author{Juan F. Mar\'in}

\affiliation{Instituto de F\'{i}sica, Pontificia Universidad Católica de Valpara\'{i}so,
Casilla 4059, Chile}

\author{Milena P\'aez-Silva}

\affiliation{Instituto de F\'{i}sica, Pontificia Universidad Católica de Valpara\'{i}so,
Casilla 4059, Chile}

\author{Majid Taki}

\affiliation{Université de Lille, CNRS, UMR 8523 - PhLAM - Physique des Lasers
Atomes et Molécules, F-59000 Lille, France.}

\author{Saliya Coulibaly}

\affiliation{Université de Lille, CNRS, UMR 8523 - PhLAM - Physique des Lasers
Atomes et Molécules, F-59000 Lille, France.}

\author{Leonardo Gordillo}

\affiliation{Departamento de Física, Universidad de Santiago de Chile~\\
Av. Ecuador 3493, Estación Central, Santiago, Chile}

\author{Mónica A. Garc\'{i}a-Ñustes}
\email{monica.garcia@pucv.cl}

\affiliation{Instituto de F\'{i}sica, Pontificia Universidad Católica de Valpara\'{i}so,
Casilla 4059, Chile}

\begin{abstract}
Faraday waves are a classic example of a system in which an extended pattern emerges under spatially uniform forcing. Motivated by
systems in which uniform excitation is not plausible, we study both experimentally and theoretically the effect of heterogeneous forcing on
Faraday waves. Our experiments show that vibrations restricted to finite regions lead to the formation of localized subharmonic wave patterns and change the onset of the instability.
The prototype model used for the theoretical calculations is the parametrically
driven and damped nonlinear Schr\"odinger equation, which is known to describe well Faraday-instability regimes. For an energy injection with a
Gaussian spatial profile, we show that the evolution of the envelope of the wave pattern can be reduced to a Weber-equation eigenvalue problem.
Our theoretical results provide very good predictions of our experimental observations provided that the decay length scale of the Gaussian
profile is much larger than the pattern wavelength.
\end{abstract}

\pacs{05.45.Yv, 05.45.-a, 89.75.Kd}

\maketitle

\section{Introduction}
\label{Sec:Intro}

Pattern formation is a major area of nonlinear dynamics \cite{nicolis1977self,RevModPhys.65.851}. During the last decades, a major progress has
been achieved in understanding
how an extended system with homogeneous conditions can spontaneously
undergo from a basic homogeneous state to a self-organized pattern \cite{cross2009pattern,pismen2006patterns}. However, a renewed interest has come from the observation of spatially localized states in uniform and non-uniform systems \cite{DewelBorckmans1989, Convectons, MoriartyHolt2011}. In uniform systems, localized patterns arise in bi-stable regions. An extended pattern solution and an homogenous one coexist, setting up a family of solutions via snaking bifurcations \cite{Snaking}. In heterogeneous media, a local spatial pattern can develop from the non-uniformity of system parameters, such as forcing or dissipation. In this latter scenario, the dynamical behavior of the system suffers modifications as corrections on the
instability domains and threshold
discretization \cite{Coulibaly2006,Huerre1990,Ouarzazi1996,PeterA.MonkewitsChomaz1993}. In particular, the concept of global mode has
been introduced to characterize the synchronized response of the system to the localization of the forcing.

Alligator's water dance is a striking example in nature of heterogenous forcing. Crocodiles and alligators are able to create spectacular local Faraday waves\textemdash spatial stationary subharmonic responses\textemdash on
the water surface through the infrasonic resonance
of their lungs \cite{Powell:2011fl,Kofron:2015wj}.  The water dance is used as an advertisement call for mating purposes of male individuals and have shown to be crucial for reproduction. Direct observations of this phenomenon, both in animals in nature or captivity, suggest that  the infrasonic radiating waves spread several kilometers under water.  Once females approach,  localized Faraday waves on the surface of the water will provide a visual signature of the size of the animal \cite{Kofron:2015wj,Dinets2013}. 

In the literature, there have been many efforts to study parametrically forced systems. Some of them have focused on the dynamics of localized structures such as solitons \cite{JunruWuRudnick1984, Gordillo2012}, local defects \cite{LeiLuLin2004,AlexeevaBarashenkov2000},
finite-size effects \cite{Clerc:2011jo,GordilloGarcia-Nustes2014},
linear-depth gradients in a water trough \cite{L.GordilloMujica2011,YanJiarenJianqiang1992}
and Gaussian parametric injection in optical systems \cite{FangweiYeMalomed2013}. However, very scarce 
studies explore the dynamics of localized Faraday waves induced by heterogenous forcing in laboratory conditions \cite{MoriartyHolt2011}. 
A thorough understanding of these systems may help us to understand how female alligators can decode the male size from Faraday waves signals. Proper tuning of heterogeneous parameters  may
also allow us to engineer the outcome of subharmonic out-of-equilibrium systems that could be used for technological applications \cite{YuJiaSeshia2013,FosterTurnerSharpingEtAl2006}.  
 
In this article, we study both experimentally and theoretically the
Faraday instability generated by localized forcing. Our experimental
setup consists of a water channel with a deformable bottom. The system
is theoretically modeled by the parametrically driven and damped non-linear Schr\"odinger (PDNLS)
equation \cite{Miles1984,MilesHenderson1990} with a spatial
varying forcing parameter. Assuming a Gaussian profile for the injection,
we use a WKBJ scaling technique \cite{Huerre1990} to derive an eigenvalue
Weber equation that governs the pattern envelope. Consequently
the response of the system is discrete and the solutions are
shown to be Hermite polynomials with Gaussian modulation.
From the
weakly nonlinear analysis, we successfully describe the nonlinear saturation of the patterns close to the threshold of the instability of the fundamental
Gauss-Hermite mode.  The theoretical results are in very good agreement with experiments.

The article is organized as follows. In section \ref{Sec:ExpSet} we show our experimental setup and describe our measurement
protocols. The theoretical description of the localized Faraday patterns are given in section \ref{Sec:Theo} and numerical simulations, in section \ref{Sec:NS}. We provide final remarks and conclusions 
in section \ref{Sec:Concl}.

\begin{figure}[ht]
\centering{}\includegraphics{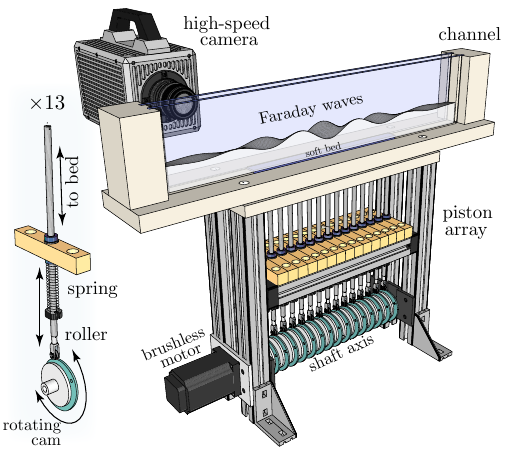}
\caption{(Color online) Experimental setup used to generate a localized injection
of energy in a Faraday-wave configuration. The soft bottom is attached
to a set of pistons, each linked to a rotary cam system attached to
a common shaft and a brushless motor. Amplitude and frequency can
be independently programmed. \label{fig01-1}}
\end{figure}

\section{Experimental setup and measurement protocol}
\label{Sec:ExpSet}

Our experimental
setup consists of a transparent rectangular water channel $15$-$\mathrm{mm}$
long, $490$-$\mathrm{mm}$ wide and $100$-$\mathrm{mm}$ deep, whose
bottom has a central soft region $240$-mm wide manufactured in a
soft silicone-elastomer (Shore hardness OO). The assembly rests over
a system of 13 pistons evenly spaced ($\Delta x=16\,\mathrm{mm}$,
each of them constrained to vertical motion by two fixed axial bearings.
At the bottom of each piston, we assembled a tiny roller to be used
as a follower. Using compressed springs, the pistons are pushed towards
a set of rotary cams placed in a common horizontal axis. The axis
is respectively coupled to a brushless motor with feedback (Model
BLM-N23-50-1000-B). 
\begin{figure*}[t]
\centering{}\includegraphics{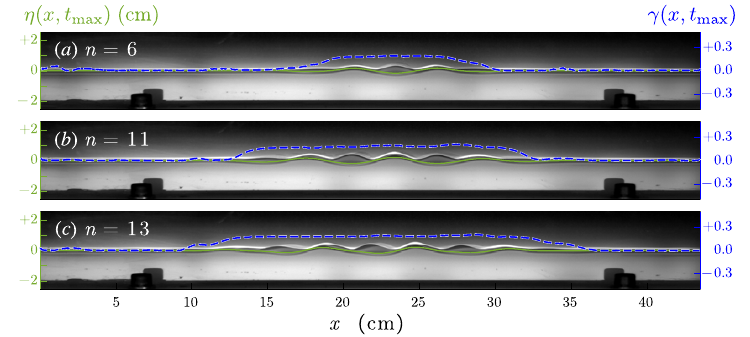}
\caption{(Color online) Snapshots of the observed structures under localized
forcing for different excitation regions: From top to bottom $\sigma_{i}=96,176,208\,\mathrm{mm}$.
Green continuous lines depict the detected free-surface deformations
at the wall, while blue dashed curves, the profiles of the effective
acceleration at the surface. \label{fig02}}
\end{figure*}
The setup resembles the mechanical transmission system of a music box as shown in Fig.~\ref{fig01-1}
and allows to deform the bottom of the channel with a spatial distribution.
Cams are shaped in such a way that an oscillatory angular motion on the axis creates
a vertical oscillatory motion on the piston. In this way, both the
acceleration amplitude $\Gamma$ (normalized by the acceleration of
gravity $g$) and frequency $f$ of oscillations can be easily programmed
through the motor controller. The motion of any piston can be easily
switched off by changing our special cams to circular ones.

The channel trough was filled with a Photoflo-water solution (concentration:
2\%) up to $20$ $\mathrm{mm}$ deep.  Under uniform forcing (e.g.
frequency $f=14.86\;\mathrm{Hz}$), Faraday waves emerge above an
acceleration threshold of $\Gamma_{c}\approx$ 0.3. The waves display
a central node in the cross-wise direction. The emerging waves were
visualized using a high-speed camera. The channel was front illuminated
so a clear vertical cut of the flow at the wall can be observed. A
small amount of white dye was added to the solution to improve visualization.

In Fig.~\ref{fig02},
we display typical images of the observed waves for different numbers
of excited pistons. The snapshots include also the reconstructed free-surface
deformation at the wall $\eta(x,t)$ and the profile of the effective
acceleration at the surface $\Gamma(x,t)$ at $t=t_{\mathrm{max}}$,
\emph{i.e. }when maximal deformations are observed through a cycle.
The curves were calculated using standard edge-detection algorithms
on the sequence of images with a sensitivity of $0.1\,\mathrm{mm}$. 

\begin{figure}[h]
\centering{}\scalebox{1.18}{\includegraphics{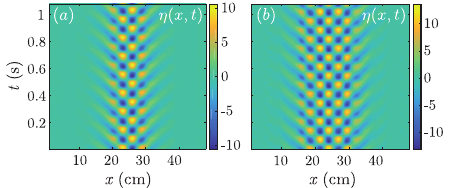}}
\caption{(Color online) Spatiotemporal evolution of free surface (Faraday waves)
at the wall $\eta(x,t)$ for (a) $n=6$ pistons and (b) $n=13$ pistons. \label{fig03}}
\end{figure}

Examples of the full spatiotemporal evolution of the free-surface
$\eta\left(x,t\right)$ are plotted in Fig.~\ref{fig03}. While obtaining
$\eta\left(x,t\right)$ is straightforward from the wavy free surface,
the effective acceleration at the surface, $\Gamma\left(x,t\right)$,
was obtained by imaging the surface of the liquid slightly below the
instability threshold $\Gamma_{c}$ where the response is still linear,
and then rescaling by the driving amplitude $\Gamma$ used in the
experiments.  It is a known issue, that the fluid layer acts as a
longpass filter of the bottom deformation so the effective deformation/acceleration
at the free surface is a smoothened version of the bottom driving
\cite{1973JFM....60..769H,Jamin:2015cn}. 

As shown in Figs.~\ref{fig02} and \ref{fig03}, the excitation of
a reduced bottom region generates wave patterns that are spatially
localized. The patterns oscillate at half the forcing frequency (parametric
instability), which is the Faraday-waves signature. The observed patterns,
that we will refer to as \emph{localized Faraday waves,} have a standing-wave
core that emits evanescent waves toward the unperturbed regions.

\subsection{ Pattern vs. injection length  }

To compare how  the wave pattern length depends on the injection length, our measurement protocol was the following: First, we chose the frequency
in such a way that the wavelength of the Faraday waves matches the
inter-piston distance. Starting from the maximum number of excited
pistons, i.e. $n=13$, we acquired a video sequence of highly resolved images
in both space and time. Then, we sequentially decreased the number
of excited pistons, hence reducing the length of the injection  region.

To characterize the wave localization, we perform the following analysis
on $\eta\left(x,t\right)$: First, we apply the temporal Fourier transform
and extract the phase and amplitude for the dominant frequency $f/2$.
In the standing-wave region, the wave displays a constant phase along
$x$ that decreases linearly as we enter into the wave-emission region.
The amplitude on the other hand, displays a smooth decay in the wave-emission
region but a serrated shape on the standing-wave. The spatial envelope
of the amplitude for the whole domain was obtained by fitting a Gaussian
curve on the amplitude local maxima of the standing-wave region and
the remaining tails of the wave-emission one. The width of the envelope
is defined as $\sigma_{w}\equiv\text{HWHM}$, i.e. the half width
at half maximum. Likewise, to characterize the injection localization,
we obtained the envelope of $\Gamma\left(x,t\right)$ and straightforwardly
obtain the $\mathrm{HWHM}$, $\sigma_{i}$. The two quantities, the
wave-envelope width $\sigma_{w}$ and the injection-envelope width
$\sigma_{i}$, were measured for runs with different number of pistons.
Results are summarized in Table \ref{tab:Exp} and displayed in Fig.~\ref{fig:04}.

\subsection{Onset of localized Faraday waves}

To study how localized Faraday waves emerge, we designed two-fold protocol for a given injection length ($n = 13$). First, we started from the flat state and increased the amplitude of oscillation with fine steps  (0.1 mm, $\Delta\Gamma = 0.086$), starting from $\Gamma_{i}= 0.265$ up to $\Gamma_{f}= 0.308$. For each given $\Gamma$, we waited  $\approx 45$ min. and checked that the wave pattern was stationary throughout several cycles before making the measurements. The second protocol was the same but we started from $\Gamma_{f}$ and decreased sequentially down to $\Gamma_{i}$.  The results are shown in Fig.~\ref{fig:09}. The system does not display hysteretic behavior, which is the signature of supercritical bifurcations.

\begin{table}[ht]
\begin{centering}
\begin{tabular}{>{\centering}p{1cm}>{\centering}p{1.5cm}>{\centering}p{1.5cm}cc}
$n$ & $\Gamma_{c}$ & $\Gamma$ & $\sigma_{i}$ (mm) & $\sigma_{w}$ (mm)\tabularnewline
\hline 
6 & 0.370 & 0.443 & 48 & 42\tabularnewline
7 & 0.345 & 0.453 & 54 & 46\tabularnewline
8 & 0.345 & 0.392 & 61 & 51\tabularnewline
9 & 0.331 & 0.382 & 68 & 57\tabularnewline
10 & 0.331 & 0.382 & 69 & 66\tabularnewline
11 & 0.331 & 0.382 & 72 & 74\tabularnewline
12 & 0.331 & 0.382 & 73 & 85\tabularnewline
13 & 0.331 & 0.382 & 80 & 92\tabularnewline
\hline 
\end{tabular}
\par\end{centering}
\caption{The table shows our experiment parameters including the forcing amplitude
$\Gamma$, the width of the injection region $\sigma_{i}$ and the width
of the wave envelope $\sigma_{w}$. In all cases, $f=14.86\,\mathrm{Hz}$.
Errors estimated from the confidence intervals of the fitted parameters
are negligible ($\lesssim0.5\,\mathrm{mm}$). \label{tab:Exp}}
\end{table}

\section{Theoretical description of localized Faraday waves}
\label{Sec:Theo}

It has been shown that the hydrodynamical problem of the free surface of a fluid which is oscillated vertically in the vicinity of Faraday instability can be reduced to an amplitude equation for 
the envelope of the surface: the PDNLS equation \cite{Miles1984,MilesHenderson1990}. Subsequently, the equation has been derived in different context as nonlinear lattices \cite{DenardoGalvinGreenfieldEtAl1992},
optical fibers \cite{KutzKathLiEtAl1993}, Kerr-type optical parametric
oscillators \cite{Longhi1996}, easy-plane ferromagnetic materials
exposed to oscillatory magnetic fields \cite{BarashenkovBogdanKorobov1991,ClercCoulibalyLaroze2012}
and parametrically driven damped chains of pendula \cite{AlexeevaBarashenkov2000}. The governing equation for the envelope of the water surface displacement of the transversal mode is	
\begin{equation}
\partial_{t}\psi=-i\nu \psi-i A |\psi|^{2}\psi-iB\partial_{x'x'}\psi-\mu \psi+\gamma\bar{\psi},
\label{Eq:PPDNLS-Dim}
\end{equation}
where $\psi(x',t)$ stands for the complex envelope of
the standing waves and $\bar{\psi}$, its complex conjugate; 
$t$ is the dimensionless time. Besides,
$\nu$ is the detuning parameter which measures the frequency offset to the
parametric resonance, $\mu$ is the damping parameter, and $\gamma$
stands for the amplitude of the parametric forcing. The parameters $A$ and $B$ are functions of the wavenumber $k$. In particular, $A\sim k ^2$ and $B \sim 1/k^2$. 
The relation between
the experiment quantities and the dimensionless parameters $\nu,\mu,\gamma$,
as well as the envelope $\psi$ and the time,  are
given in detail in Refs. \cite{Miles1984,Gordillo:2014jf, Perinet2017}. 
For our
experiments, it can be shown that $\left|\psi\right|^{2}\sim10^{-3}$,
$\mu\sim10^{-2},\nu\sim10^{-1}$ and $\gamma\sim10^{-1}$. Notice
that the PDNLS equation \eqref{Eq:PPDNLS-1} applies only in the limit
$\nu\sim\mu\sim\gamma\ll1$. For $\gamma>\mu$, $\nu>0$ and $\gamma^{2}<\nu^{2}+\mu^{2}$,
the system exhibits subharmonic patterns with critical wavelength
$k_{c}=\pm\sqrt{\nu}$, i.e., Faraday waves. Setting a dimensionless variable $x'\equiv\sqrt{B}x$ where is $x$ is the dimensionless space variable, we can rewrite the equation in the following dimensionless form,
\begin{equation}
\partial_{t}\psi=-i\nu \psi-i|\psi|^{2}\psi-i\partial_{xx}\psi-\mu \psi+\gamma\bar{\psi}.\label{Eq:PPDNLS-1}
\end{equation}

\subsection{Linear stability analysis}

We extend Eq.~\ref{Eq:PPDNLS-1} to heterogeneous systems by assuming
that $\gamma\equiv\gamma\left(x\right)$ is a function describing
the spatial profile of the forcing. 
Following our experimental results, see e.g blue dashed lines in Fig.~\ref{fig02},
we assume that the injection profile $\gamma$ is a localized function
satisfying three key features: i) $\gamma\left(x\right)$ is symmetric
with respect to a given $x_{0}$, ii) $\gamma\left(x\right)$ has a single
extremum at $x_{0}$ with non-vanishing second derivative, iii) decays
to $0$ as $x\rightarrow\pm\infty$. For the sake of simplicity, we
choose $\gamma\left(x\right)$ to be a Gaussian function:
\begin{equation}
\gamma\left(x\right)=\gamma_{i}\exp\left(-\frac{x^{2}}{2\sigma_{i}^{2}}\right),\label{Eq:Gauss}
\end{equation}
where $\gamma_{i}$ is the forcing amplitude and $\sigma_{i}$ is
a dimensionless standard deviation, $\sigma_{i}= \sigma'_{i}/\sqrt{B}$, where $\sigma'_{i}$ is a space parameter.

The degree of heterogeneity of
the system can be modeled through the parameter $\epsilon\equiv1/\sigma{}_{i}$,
which should be small to satisfy the condition of slow spatial dependence
in $\gamma\left(x\right)$. In the original physical variables, the condition $\epsilon\ll1$ is equivalent to require that
the pattern wavelength $\lambda$ is much smaller than the variation
length-scale of the forcing. Indeed, $\epsilon = 1/\sigma_{i} = \sqrt{B}/\sigma'_{i}\sim\lambda/\sigma'_{i}\ll1$, thus $\lambda\ll\sigma'_{i}$. Notice that this assumption is in agreement
with our experimental observations. 

We linearize Eq.~\ref{Eq:PPDNLS-1}
around the trivial homogeneous steady state and analyze the result and its complex conjugate (for details, see appendix \ref{App1}). 
Considering $\gamma(x)$ as a slowly varying function of space, we obtain
after some algebra, 
\begin{equation}
\left[\left(\partial_{t}+\mu\right)^{2}+\left(\nu+\partial_{xx}\right)^{2}\right]\psi-\gamma\left(x\right)\psi=0.\label{Eq:PDNLSMod}
\end{equation}
Introducing the slowly varying variable $X\equiv\epsilon x$, Eq.\eqref{Eq:PDNLSMod}
becomes
\begin{equation}
\left[\left(\partial_{t}+\mu\right)^{2}+\left(\nu+\epsilon^{2}\partial_{XX}\right)^{2}\right]\psi-\gamma^{2}\left(X\right)\psi=0,\label{Eq:PDNLSMod2}
\end{equation}
where $\gamma\left(X\right)=\gamma_{i}\left(1- X^{2}/2\right)$
is the  Taylor series expansion of $\gamma$ up to second order.

A particularly well-suited approach to find solutions in the limit
$\epsilon\ll1$ is the WKBJ approximation \cite{Orszag, DewelBorckmans1989,Huerre1990,PeterA.MonkewitsChomaz1993,Ouarzazi1996,Coulibaly2006}.
Thus,
we propose the following expansion 
\begin{equation}
\psi\left(X,t\right)=[A_{0}\left(X\right)+\epsilon A_{1}\left(X\right)+\mathcal{O}\left(\epsilon^{2}\right)]\mbox{e}^{-i\omega t}e^{\left[\frac{i}{\epsilon}\int\limits _{X_{s}}^{X}k(X)dX\right]},\label{Eq:WKBExpansion}
\end{equation}
where $A\left(X\to\infty,t,\epsilon\right)=0$, $k\left(X\right)$
is the pattern wavenumber, and $X_{s}$ is the \emph{source} point
where the heterogeneous profile is centered. Substituting the expansion
\eqref{Eq:WKBExpansion} in \eqref{Eq:PDNLSMod2}, we obtain at zeroth-order
($\epsilon^{0}$) the dispersion relation
\begin{equation}
\left(-i\omega+\mu\right)^{2}+\left(\nu-k^{2}\right)^{2}-\gamma^{2}\left(X\right)=0.\label{Eq:dispersion}
\end{equation}
Calculating higher orders of $\epsilon$ (see appendix \ref{App1}), the solution
of \eqref{Eq:PDNLSMod2} can be expressed in terms of a carrier wave
(wavelength $\sqrt{\nu}$ at dominant order) and an envelope $A_{0}\left(y\equiv\sqrt{\epsilon}x\right)$,
which obeys 

\begin{figure}[t]
\begin{centering}
\scalebox{0.5}{\includegraphics{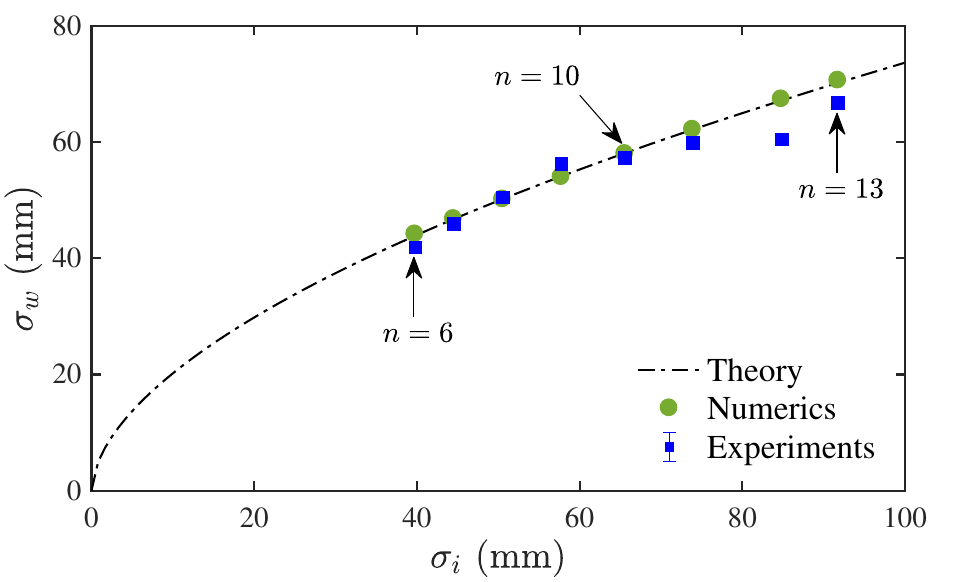}}
\par\end{centering}
\caption{Color Online: Injected forcing variance $\sigma_{i}$ as a function
of pattern variance $\sigma_{w}$ for $\nu=$ 0.07 and $\mu=$ 0.0152.
Theoretical (solid black line), experimental (blue squares) and numerical
(green circles) data are depicted. \label{fig:04}}
\end{figure}

\begin{figure}[t]
\begin{centering}
\scalebox{0.5}{\includegraphics{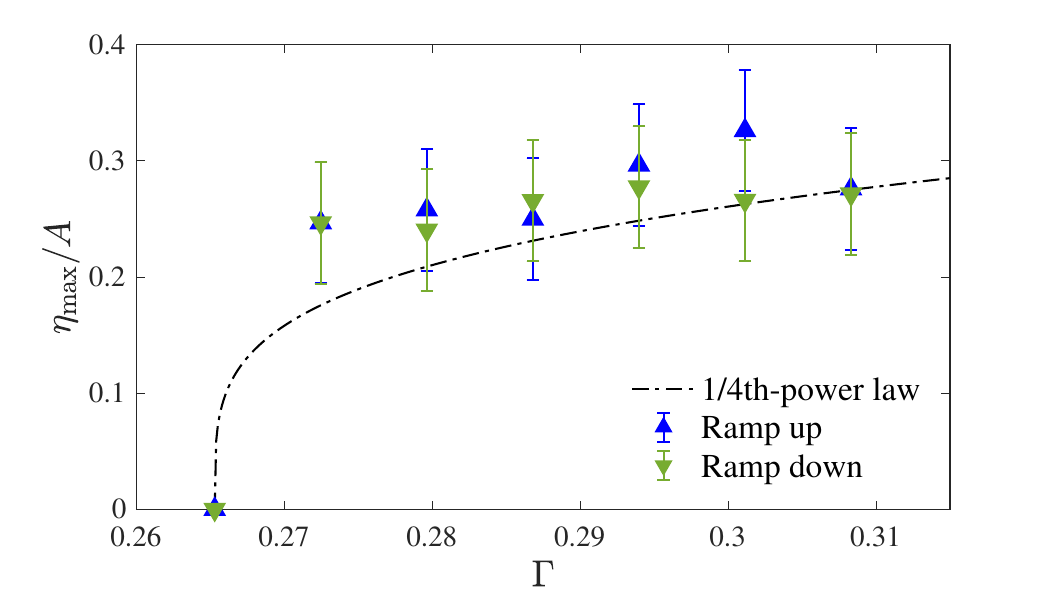}}
\par\end{centering}
\caption{Color Online: Localized Faraday-wave amplitude $\eta_{\max}$ as a function
of the forcing amplitude $\Gamma$ for $n=13$ pistons. The wave displays four nodes at frequency $f=14.61\,\mathrm{Hz}$. The two series of experimental data show increasing (blue $\blacktriangle$) and decreasing  (green $\blacktriangledown$) ramps in $\Gamma$. The data is compared with a $1/4$-th power law, $\Lambda\cdot(\Gamma - \Gamma_{c})^{1/4}$ (dashed line) with parameters $\Lambda= 0.604$ and $\Gamma_{0} = 0.265$ derived from theory (no-fitted parameters).
\label{fig:09}}
\end{figure}

\begin{equation}
\partial_{y}^{2}A_{0}+\left(\beta^{2}-\alpha y^{2}\right)A_{0}=0,\label{eq:Weber-1}
\end{equation}
where $\alpha\equiv\mu^{2}/4\nu$ and $\beta^{2}\equiv\mu\left(\gamma_{1}/2\nu+i\omega_{1}/2\nu\right)$.
Here $\gamma_{1}$ and $\omega_{1}$ have been introduced as $\gamma\simeq\mu+\epsilon\gamma^{\left(1\right)},\quad\omega\simeq\epsilon\omega^{\left(1\right)}.$
Equation~\eqref{eq:Weber-1} is a linear eigenvalue problem with
a discrete set of solutions $A_{0,m}\left(y\right)$, each given in
terms of the $m$-th Hermite polynomial $H_{m}$ modulated by a Gaussian
function; i.e. $A_{0,m}\left(y\right)=H_{m}\left(\alpha^{1/4}y\right)e^{-\left(\sqrt{\alpha}/2\right)y^{2}}$.
The eigenvalue problem also requires $\beta^{2}/\sqrt{\alpha}=2m+1$.
Further calculations show that the related quantities $\gamma_{m}^{\left(1\right)}=\left(2m+1\right)\sqrt{\nu}$
and $\omega_{m}^{\left(1\right)}=0$ are now discrete. Likewise, $\gamma_{m}=\mu+\left(2m+1\right)\left(\frac{\sqrt{\nu}}{\sigma_{i}}\right),\omega_{m}=0.$
Notice that the corrections of the mode thresholds $\gamma_{m}^{c}$ are inversely proportional to the parameter $\sigma_{i}$. This is a counterintuitive result: smaller volumes of water require higher forcing to generate patterns compared to larger ones.

We can infer that the first emerging mode in our experiments is the
fundamental one ($m=0$), which in terms of $x$ reads,
\begin{equation}
A_{0}(x)=e^{-\frac{x^{2}}{2\sigma_{w}^{2}}},\quad\mathrm{with}\ \sigma_{w}=\left(\frac{\sqrt{\nu}}{\mu}\sigma_{i}\right)^{1/2}.\label{eq:Solution}
\end{equation}
This means that all the solutions derived from \eqref{Eq:PDNLSMod2},
including the fundamental mode \eqref{eq:Solution}, display localization,
which is the key qualitative feature of localized Faraday waves. 

Equation~\eqref{eq:Solution} also shows that the width $\sigma_{w}$ of
the envelope $A_{0}$ scales as the square root of the injection-region
width $\sigma_{i}$, which agrees our experimental observation [see Fig.~\eqref{fig:04}]. Indeed, to make quantitative comparisons, we first determine the dimensionless
quantities of Eq.~\eqref{Eq:PPDNLS-1} in terms of experimental parameters.
The formulas provided in \cite{Miles1984,Gordillo:2014jf} can be
used to directly compute $\gamma_{i}$ and $\nu$. The formulas available
in the literature for the damping coefficient $\mu$ however seem
to be not appropriate for our experimental conditions since in our setup an extra shear occurs on the fixed wall. Hence we considered
$\mu$ as a phenomenological parameter, which we estimated by fitting \eqref{eq:Solution}
on the experimental data of Fig.~\ref{fig:04}. To test the validity
of our results we fitted a power law $\sigma_{w}=a\sigma_{i}^{p}$
and found $\left(a,p\right)=\left(2.27,0.480\right)$. The exponent
$p$ is remarkably consistent with the predicted square-root dependence
in \eqref{eq:Solution}.

\subsection{Weakly nonlinear analysis}

To describe the
nonlinear saturation of the unstable global modes, we have done a weakly nonlinear analysis of the system close to the spatial instability.
We introduce a bifurcation parameter $\delta\equiv\gamma_0-\gamma_0^{(1)}$ and a slowly varying amplitude $C_i(t)$ on the oscillations of
the critical mode, i.e.
\begin{equation}
\label{Eq04}
\psi_c(x,t)=C_i(t)(X_k+iY_k)e^{i\sqrt{\nu}x} + \mathrm{c.c.+ h.o.h.},
\end{equation}
where h.o.h. denotes the higher order harmonics. At the first order of nonlinearity, one obtains that $C_i$ is governed by the well known normal
form previously derived by Coullet et al. \cite{Coullet1994},
\begin{equation}
 \label{Eq05}
\partial_tC_i=\delta C_i-\frac{9}{2\mu}|C_i|^4C_i.
\end{equation}
Here, $C_i$ corresponds to the amplitude of oscillations of a single oscillator. To introduce the spatial dependence of the amplitude $C$, we first consider the dispersion relation \eqref{Eq:dispersion}, obtained from the WKBJ formalism at order $\epsilon^0$,
\begin{equation}
 \label{Eq07}
 -i\omega=-\mu\pm\sqrt{\gamma^2(X)-(\nu-k^2)^2}+\mathcal{O}(\epsilon).
\end{equation}
Assuming that the nonlinear part of Eq.~\eqref{Eq05} has slow variations in space and time compared to the linear terms, then Eq.~\eqref{Eq05}
in the Fourier space reads $-i\omega=-\mu+\gamma_0-9|C_i|^2/2\mu$. Thus, from Eq.~\eqref{Eq07} we notice that to introduce the spatial dependence in our
system we must map $\gamma_0\to[\gamma^2(X)-(\nu-k^2)^2]^{1/2}$. One obtains for the fully heterogeneous system in the Fourier space the
following expression:
\begin{equation}
\label{Eq08}
 -i\omega=-\mu+f(X,k)-\frac{9}{2\mu}|C|^4,
\end{equation}
where $f(X,k)\equiv[\gamma^2(X)-(\nu-k^2)^2]^{1/2}$. To consider the growth of modes with wavenumber $k_c=\pm\sqrt{\nu}$ due to the parametrically extended
Gaussian excitation, we consider a Taylor expansion of the function $f$ in \eqref{Eq08} for $X\equiv\epsilon x$ and $k\sim k_c=\pm\sqrt{\nu}$.
Neglecting terms of order $\mathcal{O}(\nu^{3/2}, X^4, k^4)$ and taking the inverse Fourier transform of \eqref{Eq08}, one obtains that the
amplitude $C$ of the critical mode is governed by
\begin{equation}
 \label{Eq09}
\partial_tC=\frac{2\nu}{\mu}\partial^2_xC+\left(\delta-\frac{\mu}{2\sigma^2}x^2\right)C-\frac{9}{2\mu}|C|^4C,
\end{equation}
which is a dynamical Weber-like equation with a quintic nonlinearity. In the homogeneous limit ($\sigma\to\infty$), Eq.~\eqref{Eq09} is 
similar to the normal form in \cite{Leon2014} for a spatial supercritical quintic bifurcation in a homogeneously driven
magnetic system. Notice that after taking the linear limit of \eqref{Eq09} one recovers the Weber equation \eqref{eq:Weber-1} for $\omega=0$.

To describe the nonlinear saturation of the fundamental mode, we have used a multiple-scale expansion in Eq.~\eqref{Eq09} to derive an
evolution equation for its amplitude. After some straightforward calculations\textemdash detailed in appendix \ref{App2}\textemdash one obtains that the
amplitude $D_0$ of the fundamental mode is governed by
\begin{equation}
 \label{Eq10}
 \partial_tD_0=\delta D_0-\frac{9}{2\sqrt{3}\mu}D_0^5,
\end{equation}
from which follows the stationary solution $D_0^{s}=(2\sqrt{3}\mu\delta/9)^{1/4}$. This result agrees with the experimental scaling law $\eta_{\max}\propto (\Gamma- \Gamma_{0})^{1/4}$ of Fig.~\ref{fig:09}
and describes the evolution of a supercritical quintic bifurcation. Furthermore, our theory predicts a scaling coefficient $(2\sqrt{3}\mu/9)^{1/4}= 0.482$ based on fundamental quantities ($\mu=0.14$), which is shown as a dashed line in Fig.\ref{fig:09} with no fitting parameters.

\begin{figure}[t]
\centering{}{\includegraphics[width=\columnwidth]{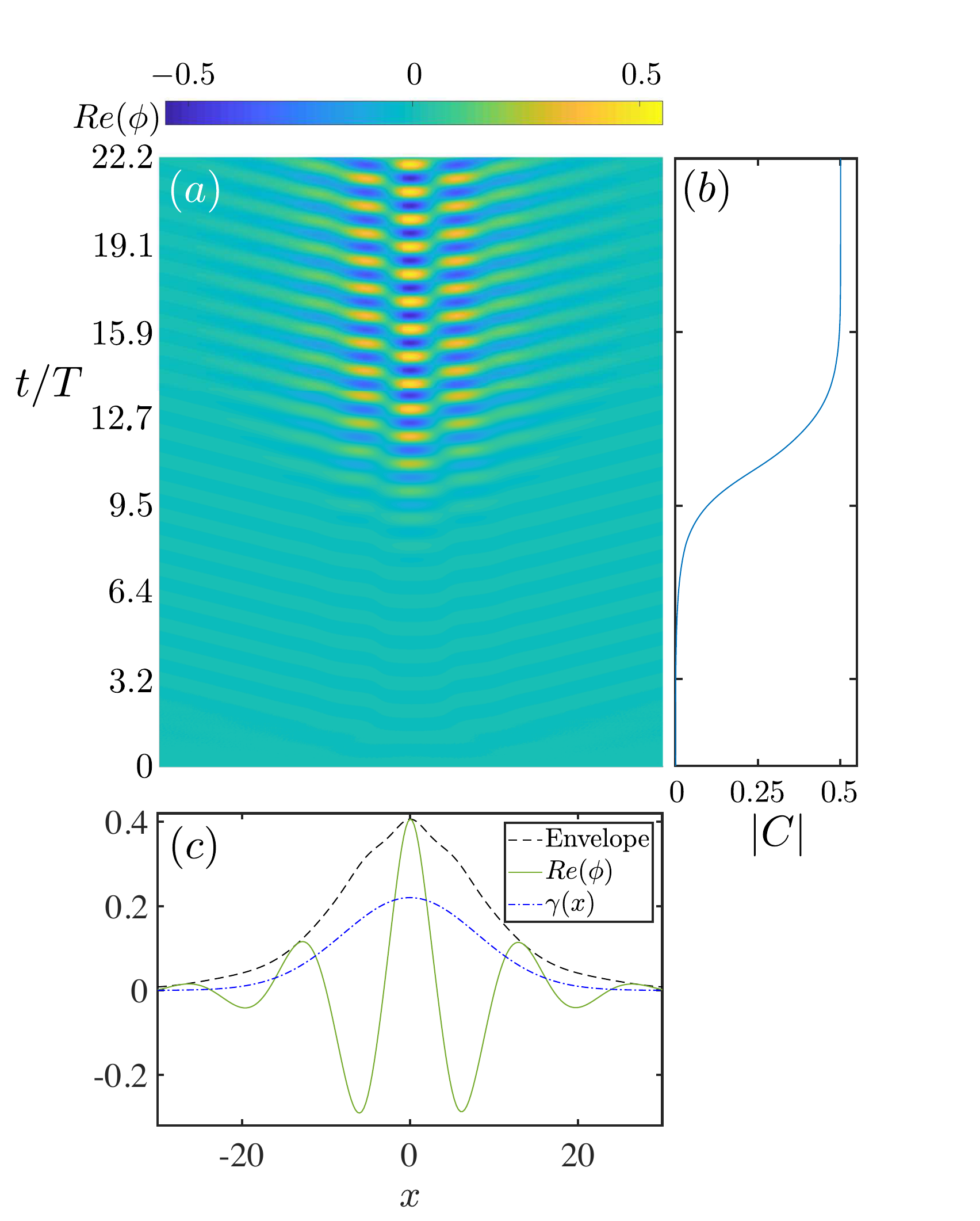}}
\caption{(Color online) Numerical simulation of the onset of a localized Faraday wave from the homogeneous zero-solution triggered by small-amplitude noise. (\textit{a}) A
wave pattern obtained with $\gamma_0=0.22$, $\nu=0.2$, $\mu=0.14$ and $\sigma=8.0$ ($\gamma_0^{(1)}<\gamma_0<\gamma_0^{(1)}$).  (\textit{b}) Temporal evolution of 
the pattern-envelope amplitude, showing the  nonlinear saturation. The envelope of the
wave at the end of the simulation has a nearly Gaussian-like profile, as shown in (\textit{c}), suggesting the instability of the fundamental
Gauss-Hermite mode. Here, the green solid line represents $\mathbb{R}e(\phi)$, while the black dashed line represents the envelope of
$\mathbb{R}e(\phi)$ obtained using the Hilbert transform. The localized injection is also depicted as a reference (dash-dotted blue line).
Time is dimensionless (measured in fundamental period units, $T$).
\label{fig06}}
\end{figure}

\section{Numerical simulations}
\label{Sec:NS}

As a final check, we also performed direct numerical simulations of Eq.~\eqref{Eq:PPDNLS-1} with no-flux boundary conditions and
$\gamma\left(x\right)$ given by the expression \eqref{Eq:Gauss}. The first goal is to determine if the linear approximation we did to obtain Eq.~\eqref{Eq:PDNLSMod2} remained valid
for the set of $\left(\mu,\nu,\gamma,\sigma_{i}\right)$ that we chose. Using a 400-point spatial grid with resolution $\mathrm{d}x=0.25$, we used
finite-differences of second order of accuracy for the space derivatives of Eq.~\eqref{Eq:PPDNLS-1}. For the time integration we ran a fourth-order Runge-Kutta scheme with a time step $\mathrm{d}t=0.0001$. 

To compare the solutions of the PDNLS equation with experiments, it is important to remark that Eq.~\eqref{Eq:PPDNLS-1} gives only a
stroboscopic evolution of the surface instability. Using $\phi\equiv\psi\exp(i\pi f t)$, where $\psi$ is the solution to the PDNLS equation \eqref{Eq:PPDNLS-1},
from $\Re e(\phi)$ we can recover the non-stroboscopic picture of the Faraday patterns. We plot the result in Fig.~\ref{fig06}(a), showing that the numerical solutions of Eq.~\eqref{Eq:PPDNLS-1} not only successfully reproduce the envelope of the localized Faraday waves but also its evanescent waves in agreement with our experiments (shown in Fig.~\ref{fig03}). 

An interesting feature that we put under test is that according to our linear stability analysis, even for $\gamma_0>\mu$, there will be no pattern formation if $\gamma_0<\gamma_0^{(0)}$. In this case,
the amplitude of the injection is greater than the dissipation but is too localized in space to sustain an instability. We have confirmed this
prediction in our experiments and numerical simulations. Indeed, any initial perturbation on the system will be eventually dissipated and end up by decaying into the homogeneous stable solution $A_h=0$.
\begin{figure*}
\centering{}\scalebox{0.4}{\includegraphics{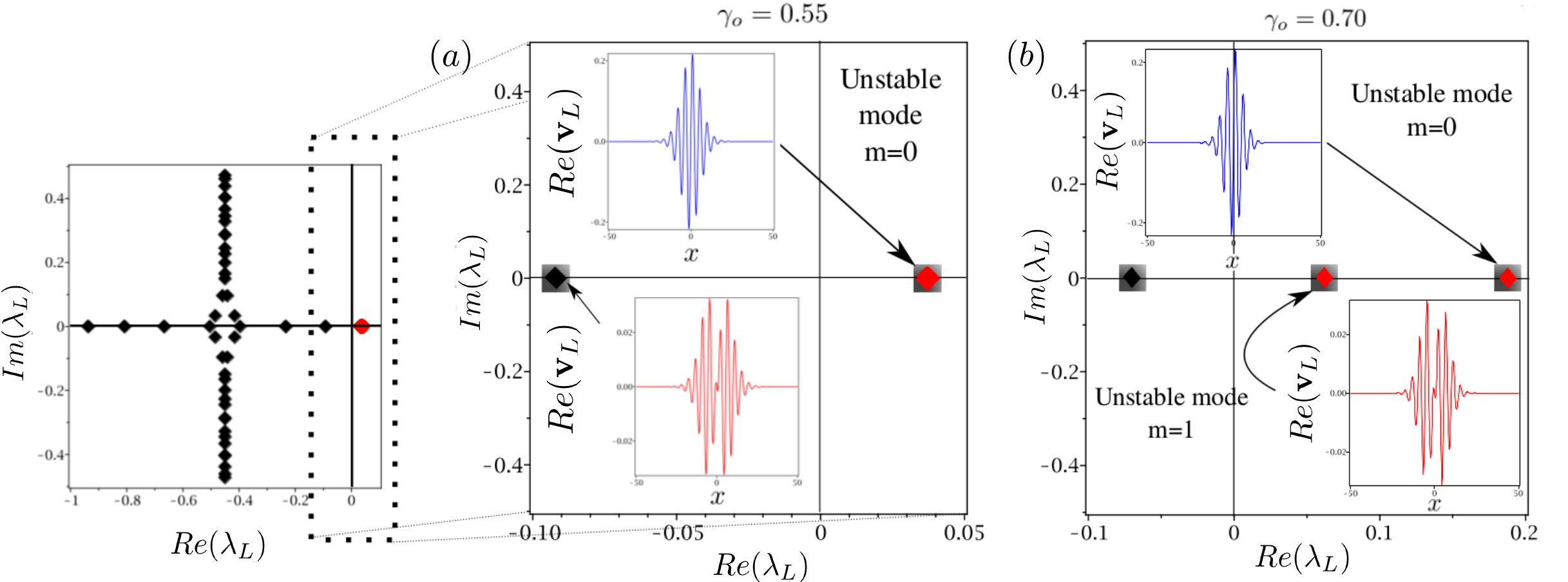}}
\caption{(Color online) Spectra obtained numerically for $\nu=1.0$, $\mu=0.45$, and $\sigma_i=16.0$. (a) Spectrum for
$\gamma_0^{(0)}<\gamma_0<\gamma_0^{(1)}$. The inset shows a zoom-out of the spectrum, showing a continuum set of eigenvalues with a non-vanishing imaginary part.
(b) Spectrum for $\gamma_0^{(1)}<\gamma_0<\gamma_0^{(2)}$. The real part of the eigenfunctions of the first two critical modes are
indicated in each case.
\label{fig07}}
\end{figure*}
However, if we increase $\gamma_0$ until we reach the region $\gamma_0^{(0)}<\gamma_0<\gamma_0^{(1)}$, a localized pattern will appear due to the
instability of the fundamental Gauss-Hermite mode. This case is shown in the numerical simulation of Fig.~\ref{fig06}(a), where we used the
homogeneous solution $A_h=0$ as the initial condition and added a small-amplitude additive noise of order $10^{-3}$. These small fluctuations are enough to trigger the Faraday instability. According to equation \eqref{Eq10}, the
amplitude of the pattern begins to grow at an exponential rate $D_0\propto\exp(\delta t)$. Nonlinear contributions become important as the amplitude of
the instability grows, and the maximum amplitude of the pattern, $|C|$, saturates due to the quintic nonlinearity of Eq.~\eqref{Eq10}, as evidenced in
Fig.~\ref{fig06}(b). 
In Fig.~\ref{fig06}(c), we show that the real part of the solution after saturation displays an envelope that has a nearly Gaussian profile.

\begin{figure}
 \centering{}\scalebox{0.35}{\includegraphics{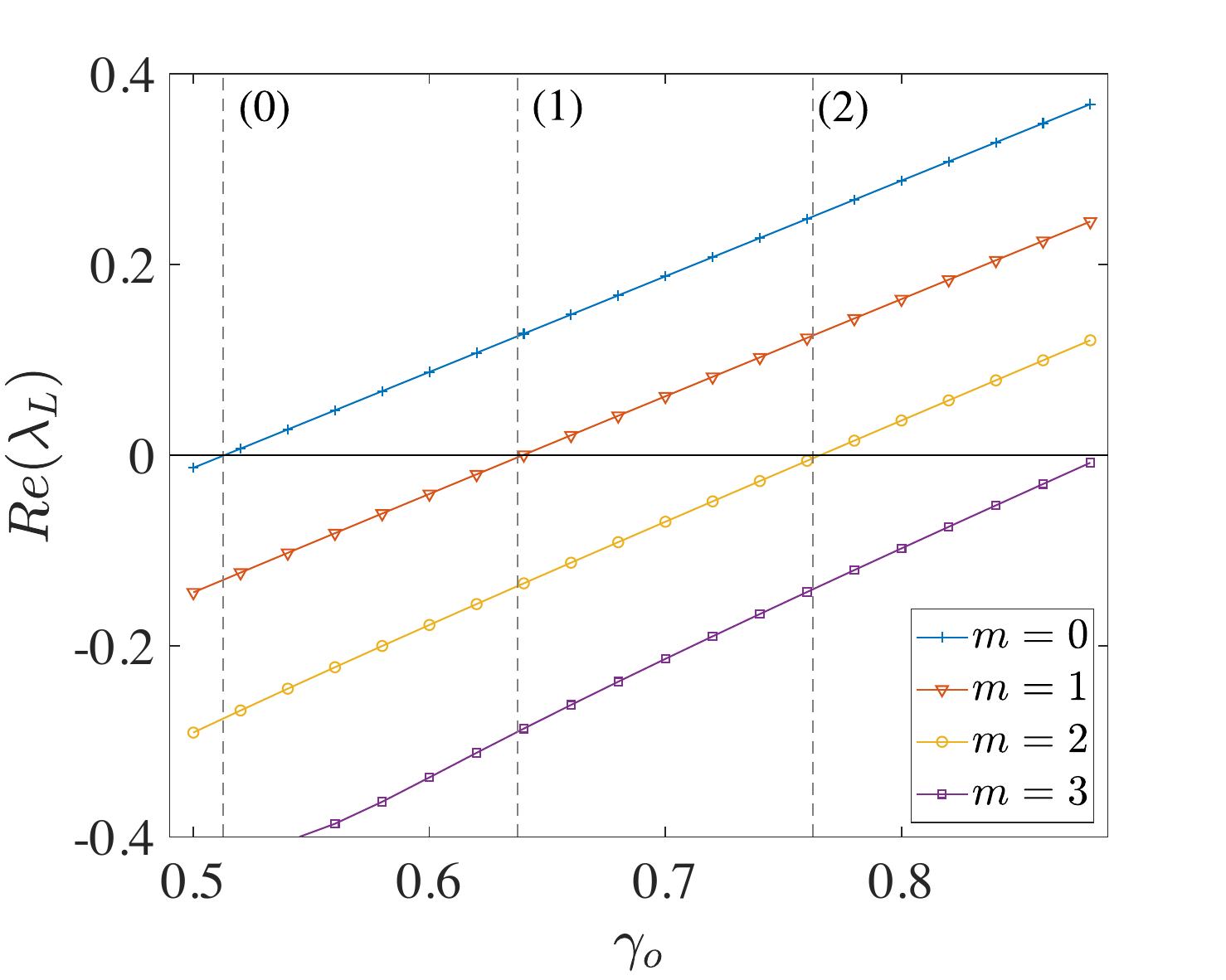}}
 \caption{(Color online) Real part of the eigenvalues of the Gauss-Hermite modes as a function of $\gamma_0$, obtained from several
numerical simulations for $\nu=1.0$, $\mu=0.45$, and $\sigma_i=16.0$. The modes turn unstable at the values predicted by the linear stability
analysis. \label{fig08}}
\end{figure}

To give a more detailed insight of the stability properties of the Gauss-Hermite modes, we have numerically computed the spectrum of
the linearized system following a similar procedure as in Ref. \cite{Clerc2010}. First, we took the numerical
solution of the PDNLS equation \eqref{Eq:PPDNLS-1} after  the envelope of the Faraday pattern has become steady. Then, we calculate the set of
eigenvalues and eigenfunctions of the linear operator that describes the dynamics of small perturbations around the Faraday-pattern solution.
Typical spectra are shown in Fig.~\ref{fig07} as well as the eigenfunctions obtained numerically for the first two eigenvalues. We expect from Eq.~\ref{Eq:PDNLSMod} that the spectrum of the linear operator is degenerate. We have confirmed this fact in our
numerical results. For $\gamma_0^{(0)}<\gamma_0<\gamma_0^{(1)}$ only the real part of the first eigenvalue is positive, as showed in
Fig.~\ref{fig07}(a). In this case, the Faraday pattern is formed due to the instability of the fundamental Gauss-Hermite mode. If we increase the
value of $\gamma_0$ until we reach the region $\gamma_0^{(1)}<\gamma_0<\gamma_0^{(2)}$, the eigenvalue of the first antisymmetric mode crosses the
imaginary axis, as shown in figure~\ref{fig07}.b. In this case, the Faraday pattern is formed due to the contributions of modes $m=0$ and $m=1$, which
are both unstable.

Finally, we have verified with several numerical simulations that the Gauss-Hermite modes turns unstable at the values of $\gamma_0$ predicted
by the linear stability analysis. Figure \ref{fig08} shows the real part of the eigenvalues of the Gauss-Hermite modes from $m=0$ to $m=3$ as a
function of the injection $\gamma_0$. The vertical dashed lines indicates the theoretical values of the thresholds of instability for each of
the modes. It is clear that each of the Gauss-Hermite modes turns unstable at the predicted values of $\gamma_0$.

In summary, the results are very consistent. The theory does not only match well the full-numerical simulation but also the
experimental results.

\section{Conclusions}
\label{Sec:Concl}

In conclusion, we have designed an experimental setup with energy injection in a spatial region whose extent can be controlled.
The setup consists  of a quasi one-dimensional rectangular
water channel with a soft deformable bottom that can be forced with a set of pistons. Above certain threshold of vibration amplitude, the water surface destabilizes into subharmonic Faraday waves that are localized in space and emit evanescent waves. 

Assuming that the width of the injection region is larger than
the characteristic pattern wavelength, we developed a WKBJ approximation and a weakly nonlinear analysis on the prototype model, i.e.,  the heterogenous PDNLS equation, to describe our experimental observations. Using this framework we have:
{\it i)} derived the spatial profile of the observed patterns; {\it ii)} showed the emission of evanescent waves; {\it iii)} computed the dependence of the envelope width on the length of the injection region; {\it iv)} described how the parameter space of the onset of instability modifies and discretizes with injection localization; {\it v)} determined that  localized Faraday waves emerge via a supercritical quintic bifurcation.  The results presented
here are helpful to understand the impact of parameter heterogeneities
on pattern-formation processes in general extended physical systems.
It is noteworthy that this work is the first complete study of localized Faraday waves in laboratory conditions and provide key  physical and mathematical insights on alligator's water dance. Further studies on localized injection with a shapeable bottom is in progress.  
 
\bigskip{}
\begin{acknowledgments}
M.A.G-N. thanks for the financial support of grant FONDECYT 11130450. L. G. was partially supported by Conicyt FCHA/Postdoctorado Becas Chile
74160007, Conicyt PAI/IAC 79160140 and FONDECYT/Iniciación 11170700. S.C., M.A.G-N., J.F.M. and M.T. thanks ECOS-Sud n\textdegree{} C15E06. J.F.M. thanks CONICYT/Doctorado Nacional 21150292. 
H. U. was partially supported by Conicyt FCHA/Beca de Doctorado en el Extranjero, Becas Chile N${}^{\circ}$ 72180269.
\end{acknowledgments}

\appendix

\section{Solutions of linearized PDNLS equation under heterogeneous forcing $\gamma(x)$}
\label{App1}

At the next order $\mathcal{O}\left(\epsilon\right)$, we obtain
the equation, $2k\left(\nu-k^{2}\right)\partial_{X}A_{0}-A_{0}\left(\nu-3k^{2}\right)\partial_{X}k=0$.
Using relation \eqref{Eq:dispersion}, we can now deduce an expression
for $A_{0}$, i.e.
\begin{equation}
A_{0}\propto\mbox{exp}\left[\frac{-i\int F(k,\omega,\mu,\gamma_{0},\nu,X)\,\mathrm{d}X}{\partial\omega/\partial k}\right],\label{Eq:diverges}
\end{equation}
where $F(k,\omega,\mu,\gamma_{0},\nu,X)\equiv\frac{\nu-3k^{2}}{-i\omega+\mu}\partial_{X}k$.
It is clear that for $\partial\omega/\partial k=0$, expression \eqref{Eq:diverges}
is singular. The points where the singularity $\partial\omega/\partial k=0$
take place are called turning points \cite{Orszag}. On these points,
the WKBJ approach is not longer valid. Let be $\omega_{0}\equiv\omega\left(k_{0}\right)$,
where $k_{0}$ is such that 
\begin{equation}
\frac{\partial\omega\left(X\right)}{\partial k}\big|_{k=k_{0}}=0.
\end{equation}
By definition, if the contour in the complex plane $\mbox{Im}[\omega_{0}\left(X\right)]$
exhibits a contact line or pinch between two branches $X^{\pm}(\omega_{0})$
in a turning point $X^{t}$ that also verifies $\partial\omega_{0}\left(X^{t}\right)/\partial_{X}=0$,
then we deal with a double turning point \cite{Huerre1990,PeterA.MonkewitsChomaz1993,DewelBorckmans1989,Ouarzazi1996,Coulibaly2006}.
In that case, the scaling law that rules the system dynamics in the
region close to the double turning point is $X=\epsilon^{1/2}y$.
In consequence, the expansion takes the following new form, 
\begin{equation}
\psi\left(y,t\right)=[A_{0}\left(y\right)+\epsilon^{1/2}A_{1}\left(y\right)+\mathcal{O}\left(\epsilon\right)]e^{-i\omega t}e^{\left[\frac{i}{\epsilon^{1/2}}\int\limits _{y_{s}}^{y}k(y)\,\mathrm{d}y\right]}.\label{Eq:WKBExpansion2}
\end{equation}
As in the homogeneous case, we have two solutions for $\partial\omega/\partial k=0$,
i) $k_{c}=0$ and ii) $k_{c}=\pm\sqrt{\nu}$ with $\nu>0$, which
correspond respectively to the angular frequencies 
\begin{align}
\omega(k_{c}) & =i\left(-\mu+\sqrt{\gamma^{2}\left(X\right)-\nu^{2}}\,\right)\quad\mbox{and}\\
\omega(k_{c}) & =i\left(-\mu+\gamma\left(X\right)\right).
\end{align}
Imposing the second condition ($\partial\omega/\partial X=0$) for
the double turning point, it follows that $X^{t}=0$ for both cases.
On the contrary, for different $k_{c}$, we have different values
of $\gamma^{t}$. For Faraday waves ($\nu>0$, $\gamma>\mu$), the
critical wavelength is $k_{c}=\pm\sqrt{\nu}$ and $\gamma^{t}=\mu$.


After making the corresponding replacements in the parameter expansions
and provided that $\epsilon\ll1$, the spatial forcing takes the
form $\gamma\left(x\right)=\gamma_{i}(1-\epsilon\left(y-y^{t}\right){}^{2}/2+\mathcal{O}\left(\epsilon^{2}\right)$.
We hence introduce small deviations from the turning point $\left\{ X^{t},\gamma^{t}\right\} $
through 
\begin{align}
\gamma & =\gamma^{t}+\epsilon\gamma^{\left(1\right)}+\mathcal{O}\left(\epsilon^{2}\right),\\
\omega & =\omega^{t}+\epsilon\omega^{\left(1\right)}+\mathcal{O}\left(\epsilon^{2}\right).
\end{align}
Thus, we get the expression for the forcing in the turning point,
$\gamma\left(y\right)=\gamma^{t}+\epsilon\left[\gamma^{\left(1\right)}-\gamma_{0}^{t}\left(y-y^{t}\right){}^{2}/2\right]+\mathcal{O}\left(\epsilon^{2}\right)$.
At dominant order. $k=k_{c}=\sqrt{\nu}$ and Eq.~\eqref{Eq:WKBExpansion2}
reduces to $\psi\left(y,t\right)\sim\mbox{{exp}}\left[i\int\sqrt{\nu}\,\mathrm{d}y/\epsilon^{1/2}\right]$
\cite{DewelBorckmans1989}. Next, we replace $\psi\left(y,t\right)$
in \eqref{Eq:PDNLSMod} and analyze the equations
in orders of $\epsilon$. At order $\mathcal{O}\left(\epsilon^{0}\right)$,
we obtain that $A_{0}=\bar{A_{0}}$. At order $\mathcal{O}\left(\epsilon^{1/2}\right)$,
we get the relation 

\begin{equation}
\mu\left(A_{1}-\bar{A}_{1}\right)=2\sqrt{\nu}\partial_{y}A_{0}.\label{eq:Order1/2}
\end{equation}
Finally at order $\mathcal{O}\left(\epsilon\right)$, using $A_{0}=\bar{A_{0}}$
and relation Eq.~\eqref{eq:Order1/2}, we obtain a \textit{Weber
equation} that describes the linear behavior of the signal envelope
$A_{0}$, 
\begin{equation}
\partial_{y}^{2}A_{0}+\left(\beta^{2}-\alpha y^{2}\right)A_{0}=0\label{eq:Weber}
\end{equation}
where $\text{\ensuremath{\alpha}}\equiv\mu^{2}/4\nu$ and $\beta^{2}\equiv\mu\left(\gamma^{\left(1\right)}/2\nu+i\omega^{\left(1\right)}/2\nu\right)$.
The solutions of \eqref{eq:Weber} are Hermite polynomials with a
Gaussian modulation, i.e. $A_{0}=H_{m}\left(\alpha^{1/4}y\right)e^{-\left(\sqrt{\alpha}/2\right)y^{2}}$.
Due to the discrete spectrum of the linear operator in \eqref{eq:Weber},
we also require $\beta^{2}/\sqrt{\alpha}=2m+1$, which imposes conditions
over $\gamma^{\left(1\right)}$ and $\omega^{\left(1\right)}$. In
terms of the original variable $x$, the solution is,

\[
A_{0}\left(x\right)=H_{\text{\ensuremath{\nu}}}\left(\alpha^{\frac{1}{4}}\sigma_{i}^{-\frac{1}{2}}x\right)e^{-x^{2}/2\sigma_{w}^{2}}
\]
where $H_{m}$ are the Hermite polynomials and the standard deviation
$\sigma_{w}$ is given as 
\[
\sigma_{w}^{2}\equiv\left(\frac{1}{\sqrt{\alpha}}\right)\sigma_{i}=\left(\frac{2\sqrt{\nu}}{\mu}\right)\sigma_{i}.
\]
We have shown that a weak spatial dependence of $\gamma$ generates
a modulation on the wave pattern given by an amplitude equation equivalent
to a Weber-equation eigenvalue problem. 

\section{Nonlinear saturation of the fundamental mode}
\label{App2}

In this appendix, we give the derivation of the evolution equation for the amplitude of the fundamental Gauss-Hermite mode, when the system is
close to the threshold of instability $\gamma_0^c$. Let us define $ \delta_{m}^c:=\gamma_m^c-\mu$, where
$\gamma_m^c=\mu+\sqrt{\nu}(2m+1)/\sigma_i$. We
perform a multiscale development \cite{Peyrard2004} in eq.~\eqref{Eq09} introducing new variables with different
time-scales according to $T_i:=\varepsilon^{i} t$ (with $i=1,2,\dots$). Here, $\varepsilon$ is a small adimensional parameter introduced only for
the multiscale analysis. We search the field $C(x,t)$ in the form of a perturbative development of functions of the different time-scales
according to
\begin{equation}
 C(x,t):=\sum_{i=1}^{\infty}\varepsilon^{i/4}A_i(x, T_1, T_2, \ldots),
\end{equation}
with a similar expansion in the bifurcation parameter
\begin{eqnarray}
  \delta:=\delta_0+\varepsilon\delta_1+\varepsilon^2\delta_2+\varepsilon^3\delta_3+\mathcal{O}(\varepsilon^4).
\end{eqnarray}

Considering all these developments in eq.~\eqref{Eq09}, we proceed to analyse the system at each order of $\varepsilon$.

\subsection{Order $\varepsilon^{1/4}$}

At order $\varepsilon^{1/4}$, Eq.~\eqref{Eq09} reads
\begin{equation}
\label{AppB:Eq01}
 \partial^2_xA_1+\frac{\mu}{2\nu}\left(\delta_m^c-\frac{\mu}{2\sigma^2}x^2\right)A_1=0,
\end{equation}
which is the Weber equation. Solutions of Eq.~\eqref{AppB:Eq01} are given by the Gauss-Hermite polynomials, which are denoted here as
$A_1^{(n)}(x)\equiv H_n(\alpha^{1/4}x/\sigma_i^{1/2})\exp(-x^2/2\sigma_w^2)$.
Thus, the general solution of Eq.~\eqref{AppB:Eq01} can be written as
\begin{equation}
\label{AppB:Eq02}
A_1=D_0(T_1,\ldots)A_1^{(0)}+\sum_{n=1}^{\infty}D_n(T_1,\ldots)A_1^{(n)},
\end{equation}
which is a linear combination of the Gauss-Hermite modes with time-dependent coefficients. The fundamental mode is given by
\begin{equation}
\label{AppB:Eq03}
 A_1^{(0)}(x)=e^{-x^2/2\sigma_w^2},
\end{equation}
where $\sigma_w^2\equiv\sigma_i\sqrt{\nu}/\mu$. Thus, the saturation of the fundamental mode \eqref{AppB:Eq03} will be given by the time evolution of the
coefficient $D_0(T_1,\ldots)$ in eq.~\eqref{AppB:Eq02}.

\subsection{Order $\varepsilon^{5/4}$}

At order $\varepsilon^{5/4}$, Eq.~\eqref{Eq09} can be written as the linear problem
\begin{equation}
\label{EqLinearProb}
LA_5=b, 
\end{equation}
where
\begin{eqnarray}
L\equiv \partial_x^2+\frac{\mu}{2\nu}\left(\delta_0-\frac{\mu x^2}{2\sigma^2}\right),\\
\label{b}
b\equiv \partial_{T_1}A_1-\delta_1A_1+\frac{9}{2\mu}A_1^5.
\end{eqnarray}
The linear problem \eqref{EqLinearProb} can be solved for $A_5$ only if $b$ is in the image of the operator $L$. According to the Fredholm
alternative \cite{pismen2006patterns}, at least one solution for $A_5$ exists if $\exists!|v\rangle\in\ker L^{\dagger}$ such that
$\langle v| b\rangle=0$. Notice that $A_1^{(0)}\in\ker L^{\dagger}$, since $LA_1^{(0)}=0$ and $L=L^{\dagger}$. Thus, the Fredholm alternative
gives
\begin{equation}
 \label{AppB:Eq04}
\langle A_1^{(0)}| b\rangle=0.
\end{equation}
The modes $A_1^{(n)}$ for $n\geq2$ are all stable near the threshold of instability of the fundamental mode. Thus, the amplitudes
$D_n(T_1)$ for $n\geq2$ decays exponentially in time. Once the pattern has completely evolved, one simply obtains
\begin{equation}
 \label{AppB:Eq05a}
 A_1\simeq D_0(T_1)A_1^{(0)},
\end{equation}
which is an even function in space. Inserting Eq.~\eqref{AppB:Eq05a} in equations \eqref{b} and \eqref{AppB:Eq04}, one obtains the solvability
condition
\begin{equation}
 \label{AppB:Eq06}
 \partial_{T_1}D_0=\delta_1D_0-\frac{9}{2\sqrt{3}\mu}D_0^5,
\end{equation}
from which follows Eq.~\eqref{Eq10}.

\end{document}